\documentclass[article]{ws-ijmpa}
\usepackage[super,compress]{cite}
\usepackage{graphicx}
\begin{document}
\markboth{Ayan Mukhopadhyay}{The holographic principle and RG flow}

\title{Understanding the holographic principle via RG flow}

\author{Ayan Mukhopadhyay}

\address{Institut f\"ur Theoretische Physik, Technische Universit\"at Wien,
        Wiedner Hauptstrasse 8-10, A-1040 Vienna, Austria and\\
        CERN, Theoretical Physics Department, 1211 Geneva 23, Switzerland\\
ayan@hep.itp.tuwien.ac.at}

\maketitle


\begin{abstract}
This is a review of some recent works which demonstrate how the classical equations of gravity in AdS themselves hold the key to understanding their holographic origin in the form of a strongly coupled large $N$ QFT whose algebra of local operators can be generated by a few (single-trace) elements. I discuss how this can be realised by reformulating Einstein's equations in AdS in the form of a non-perturbative RG flow that further leads to a new approach towards constructing strongly interacting QFTs. In particular, the RG flow can self-determine the UV data that are otherwise obtained by solving classical gravity equations and demanding that the solutions do not have naked singularities. For a concrete demonstration, I focus on the hydrodynamic limit in which case this RG flow connects the AdS/CFT correspondence with the membrane paradigm, and also reproduces the known values of the dual QFT transport coefficients. 
\end{abstract}

\ccode{PACS numbers: 11.25.Tq, 11.10.Gh, 04.20.Dw}

\section{A route to explore the holographic origin of gravity}
\label{sec:1}

It is widely believed that \textit{the holographic principle} holds the key to merging quantum and gravity together into a consistent framework. This principle broadly postulates that the gravitational dynamics in a given \textit{volume} of spacetime can be described using degrees of freedom living at the \textit{boundary} \cite{tHooft:1999bw,Susskind:1994vu,Bousso:2002ju,Maldacena:2003nj}. Thus gravity and at least one dimension of spacetime should be \textit{both} emergent together from familiar quantum dynamics of many-body systems living on a holographic screen, whose embedding in the emergent spacetime should depend on the observer and the measurement process. A precise general statement of the holographic principle is still elusive although we do have a very concrete realisation in the form of AdS/CFT correspondence of string theory in which certain supergravity theories with stringy corrections in anti-de Sitter (AdS) space have been shown to have  \textit{dual} descriptions given by specific types of conformal Yang-Mills theories without gravity living at the boundary \cite{Maldacena:1997re,Witten:1998qj,Gubser:1998bc}.

Recently, a specific approach towards understanding of the holographic principle  (specially \cite{Kuperstein:2011fn,Kuperstein:2013hqa,Behr:2015yna,Behr:2015aat}) has been developed which has been directly influenced by the broad philosophy that the classical gravity equations themselves hold the key to unravelling gravity's holographic origin. In the context of AdS/CFT correspondence, which is the most concrete example of holography, this question can be formulated in a precise way. Let us however state this question from a broader point of view by considering a class of gravitational spacetimes where one can naturally define a \textit{spatial holographic direction} related with a \textit{decreasing} energy scale. Such spacetimes include asymptotically anti-de Sitter (aAdS) spaces and those with horizons (such as black holes) where this holographic direction is the radial direction associated with a \textit{wrap factor} or a \textit{blackening function}. \textit{If gravity is holographic, then the holographic radial direction should be related to a scale of precise kind of renormalisation group flow of the dual quantum system implying that holographic screens at constant values of this radial coordinate should contain complete information about a specific kind of coarse-grained description of the dual quantum system.} The broad question partly is, how do we make this statement precise and also how do we relate the freedom of choice of local coarse-graining for doing measurements in the dual quantum system to the emergence  of diffeomorphism symmetry in the gravitational theory. 

In the case of AdS/CFT correspondence, the precise microscopic QFT described by the data on the holographic screen at infinity (the boundary of the aAdS spacetime) is precisely known. Nevertheless, the precise general relation between the scale of the QFT and the emergent radial coordinate, meaning a correspondence between RG flow in the QFT and the radial evolution of data on holographic screens via gravitational dynamics is still unknown. In this article, we will discuss how an amazing lot about this mysterious map between RG flow and radial evolution can be known by an appropriate reformulation of classical gravity equations themselves.

There is another aspect of the holographic origin of gravity which is also very enigmatic. Typically the map between classical gravity with a few fields and a dual QFT works only when the latter is strongly coupled \cite{Maldacena:1997re,Witten:1998qj,Gubser:1998bc}. This feature has revolutionised our understanding of strong coupling dynamics in quantum many-body systems. At strong coupling, the perturbative machinery of calculations with Feynman diagrams does not work and so far there is no better alternative to the holographic duality (whenever it is applicable) for calculating real-time quantum dynamics in presence of strong interactions. In order to calculate physical quantities via holography, one simply solves for the \textit{asymptotic} data that lead to solutions in the dual gravity theory which are \textit{free of naked singularities}. These lead to relations between the apriori independent leading (non-normalisable) and subleading (normalisable) modes of the gravitational fields near the boundary of AdS which satisfy two-derivative equations (such as Einstein's equations and the covariant Klein-Gordon equation). Each such field corresponds to an operator of the quantum theory. The non-normalisable modes correspond to the \textit{sources} for local operators, and the normalisable modes correspond to the \textit{expectation values} of the corresponding operators. Solving for the relations between these two that lead to dual geometries without naked singularities, we can obtain correlation functions, transport coefficients, etc. of the dual QFT. In fact, even if the Lagrangian description of the dual QFT is unknown, the dual gravity description gives us a concrete machinery to calculate all physical observables.

The enigmatic aspect is as follows. If the classical gravity equations can be reformulated as a RG flow, then this RG flow itself should know which microscopic UV data should lead to dual spacetimes in the theory of gravity without naked singularities. The RG flow is a first order evolution with the holographic radial direction moving towards the infrared of the dual QFT. The criterion for absence of naked singularities should be better obtained from the infrared behaviour of the RG flow, as often  the dual field theory can become weakly coupled in the ultraviolet so that the holographic classical gravitational description may no longer be valid \cite{Klebanov:2000hb}. Therefore, demanding appropriate requirements on the infrared holographic screen where the RG flow ends should ensure absence of naked singularities in dual gravity. Typically, this infrared holographic screen is the horizon. However, this infrared horizon screen should not be a fixed point of the RG flow so that the microscopic UV data can be recovered from the endpoint data by following back the first order scale evolution. \textit{The question then is what is this infrared behaviour of RG flow that should be specified at the endpoint (the holographic screen coinciding with the horizon) which should lead us to the same UV data that is usually specified at the AdS boundary to ensure that the dual spacetimes do not have naked singularities.}

The data at the holographic horizon screen is expected to be very universal and characterised by a few parameters. As for example, although the microscopic UV data in the hydrodynamic limit consists of infinite number of transport coefficients which should be specified at the AdS boundary to obtain regular future horizons \cite{Policastro:2001yc,Baier:2007ix,Bhattacharyya:2008jc}, the dynamics of the horizon is known to be characterised universally by a \textit{non-relativistic incompressible Navier-Stokes fluid} with the shear viscosity being the only parameter as demonstrated via the membrane paradigm \cite{Damour:1978cg,1986bhmp.book.....T}. Therefore, somehow the endpoint of RG flow that reformulates classical gravitational dynamics should be specified only by a \textit{few} parameters which should determine the infinite number of physical observables of the dual QFT.  A natural implication then is that \textit{in any fixed number of dimensions only a class of gravitational theories (which may be constituted by finite or infinite number of higher derivative corrections to Einstein's equations) can be holographic -- such gravitational theories can be parametrised by a finite number of infrared parameters}. Furthermore, \textit{this can possibly be revealed by reformulating the classical equations of gravity in the form of RG flows, and then finding out when the absence of naked singularities in solutions of the gravitational theory can be translated into an appropriate criterion for the endpoint of the RG flow which can be described by simple dynamical equations involving a few parameters only.}

In the following section, I will describe how such a reformulation of classical gravity equations in AdS in the form of  RG flows work and also how the infrared criterion for the RG flow can determine the microscopic UV data of the dual field theories. In Section 3, I will describe the construction of the RG flow in the field theory which will also define the latter in a constructive way in special limits. Special emphasis will be given on the hydrodynamic sector. I will conclude with an outlook.

\section{Reformulating gravity as a \textit{highly efficient RG flow}}
The map between gravity in AdS and RG flow can be readily understood in the Fefferman-Graham coordinates which is well adapted for the description of the asymptotic behaviour of the spacetime metric and other gravitational fields from which the microscopic UV data of the dual field theory can be readily extracted. Therefore, we first describe how the map works in the Fefferman-Graham coordinates. The map can of course be expressed in any coordinate system and this will be related to the freedom of choosing the scale of observation in the dual field theory locally as we will show later. We will also consider pure Einstein's gravity for most of our discussion.

Any asymptotically AdS (aAdS) spacetime metric can be expressed in the Fefferman-Graham coordinates in the form:
\begin{equation}\label{FG}
{\rm d}s^2 = \frac{l^2}{r^2}\left({\rm d}r^2 + g_{\mu\nu}(r,x){\rm d}x^\mu {\rm d}x^\nu \right).
\end{equation}
This coordinate system should be valid in a finite patch ending at the boundary which is at $r= 0$. Also $l$ is called the AdS radius and in the holographic correspondence it provides units of measurement of bulk gravitational quantities which then corresponds to parameters and couplings of the dual field theory. The \textit{boundary metric} $g^{\rm (b)}_{\mu\nu}$ defined as:
\begin{equation}
g^{\rm (b)}_{\mu\nu}(x) \equiv g_{\mu\nu}(r = 0, x)  
\end{equation}
is identified with the metric on which the dual field theory lives. For the sake of simplicity, unless stated otherwise we will assume that $g^{\rm (b)}_{\mu\nu} = \eta_{\mu\nu}$ so that the dual field theory lives in flat Minkowski space. Our results of course can be generalised readily to any arbitrary curved boundary metric.

For later purposes, it is useful to define:
\begin{equation}
z^\mu_{\phantom{\mu}\nu} \equiv g^{\mu\rho}\frac{\partial}{\partial r}g_{\rho\nu}.
\end{equation}
Einstein's equations with a negative cosmological constant $\Lambda = -d(d-1)/(2l^2)$ in $(d+1)-$ dimensions in Fefferman-Graham coordinates can be written in the following form \cite{Kuperstein:2013hqa}:
\begin{eqnarray}\label{Einstein}
\frac{\partial}{\partial r}{z^\mu_{\phantom{\mu}\nu}} - \frac{d-1}{r} z^\mu_{\phantom{\mu}\nu} + {\rm Tr}\, z\left(\frac{1}{2}z^\mu_{\phantom{\mu}\nu}-\frac{1}{r}\delta^\mu_{\phantom{\mu}\nu}\right)
&=& 2 \, R^\mu_{\phantom{\mu}\nu},\label{tensor-equation}\nonumber\\
\nabla_\mu\left(z^\mu_{\phantom{\mu}\nu} -{\rm Tr}\, z\delta^\mu_{\phantom{\mu}\nu}\right)&=& 0, \label{vector-equation}\nonumber\\
 \frac{\partial}{\partial r}{{\rm Tr}\, z} -\frac{1}{r}{\rm Tr}\, z +\frac{1}{2}{\rm Tr}\,z^2 &=& 0.
\end{eqnarray}
Above, all indices have been lowered or raised with $g_{\mu\nu}$ or its inverse respectively. The first equation is the real dynamical equation and the latter ones are constraints that the data at that boundary $r=0$ should satisfy. The radial dynamical evolution preserves the constraints, and therefore if they are satisfied at $r=0$, they should be satisfied everywhere (for further details see \cite{Gupta:2008th}). It is to be noticed that in the above form the AdS radius $l$ does not appear in the equations of motion.

Let us proceed now \textit{without} assuming the traditional rules of AdS/CFT correspondence. We only assume that corresponding to any \textit{solution} of $(d+1)-$ dimensional Einstein's equations with a negative cosmological constant that is free of naked singularities, there should exist a \textit{state} in the dual $d-$dimensional field theory. The $(d+1)$ metric then should contain information about $\langle {t_{\mu\nu}}^\infty \rangle$, the expectation value of the microscopic energy-momentum tensor operator in the dual QFT state. For later convenience of analysing the hydrodynamic limit, we will consider $\langle {t^\mu_{\phantom{\mu}\nu}}^\infty \rangle$ instead of $\langle {t_{\mu\nu}}^\infty \rangle$. When the boundary metric is $\eta_{\mu\nu}$, $\langle {t^\mu_{\phantom{\mu}\nu}}^\infty \rangle$ should satisfy the Ward identities:
\begin{equation}\label{WIinfty}
\partial_\mu \langle {t^\mu_{\phantom{\mu}\nu}}^\infty \rangle = 0, \quad {\rm Tr}\, \langle t^\infty \rangle = 0
\end{equation}
The first one is the local conservation of the energy-momentum tensor and the latter comes from conformal invariance (we will later see why the dual quantum theory should have conformal invariance). The question is of course how to extract $\langle {t^\mu_{\phantom{\mu}\nu}}^\infty \rangle$ from the dual spacetime metric. At this stage, it should be intuitively obvious that the microscopic Ward identities (\ref{WIinfty}) should be related to the constraints of Einstein's equations (\ref{Einstein}).

We should now see the problem of identification of $\langle {t^\mu_{\phantom{\mu}\nu}}^\infty \rangle$ from a broader perspective of connecting data on holographic screens at $r = \textit{constant}$ with an appropriate RG flow in the dual QFT. Firstly, we identify the radial coordinate $r$ with the inverse of the scale $\Lambda$ of the dual quantum theory, i.e. we impose the relation
\begin{equation}
r = \Lambda^{-1}.
\end{equation}
If the relation between $r$ and $\Lambda$ should be such that (i) it is state (i.e. solution) independent, and (ii) that the AdS radius $l$ which has no direct interpretation in the dual QFT should not play a role in the mutual identifications, then the above is the only possibility given that $r=0$ corresponds to the UV. Now on holographic screens at $r = \textit{constant}$ we must identify the following pair of data $g_{\mu\nu}(\Lambda)$ and $\langle {t^\mu_{\phantom{\mu}\nu}}(\Lambda) \rangle$. The effective metric $g_{\mu\nu}(\Lambda)$ can be seen as a generalised effective scale-dependent coupling or rather the source for the effective operator $\langle {t^\mu_{\phantom{\mu}\nu}}(\Lambda) \rangle$. We should identify $g_{\mu\nu}(\Lambda)$ with $g_{\mu\nu}(r)$ that appears in the Fefferman-Graham metric (\ref{FG}) at $r=\Lambda^{-1}$ for reasons similar to those mentioned above. Firstly, as evident from (\ref{tensor-equation}), as a result of this identification the evolution equations for $g_{\mu\nu}(\Lambda)$ does not involve $l$ which has no direct meaning in the dual QFT, and secondly the identification is also state (solution) independent. Furthermore, $g_{\mu\nu}(\Lambda)$ coincides then with the metric $\eta_{\mu\nu}$ on which the dual QFT lives at $\Lambda = \infty$. In usual perturbative RG flows, we do not talk about a background metric $g_{\mu\nu}(\Lambda)$ that evolves with the scale, however it makes perfect sense to do so in a special limit as explained below.

At this stage, we can introduce the notion of \textit{highly efficient RG flow} \cite{Behr:2015yna,Behr:2015aat}. To understand this notion, it is first useful to classify operators in a QFT as \textit{single-trace} and \textit{multi-trace} operators. Single-trace operators are those which are gauge-invariant and which form the minimal set of generators of the algebra of all local gauge-invariant operators. All other gauge-invariant operators, which are multi-trace, are formed out of products of the single-trace operators and their spacetime derivatives. It is these single-trace operators which are dual to gravitational fields in the holographic correspondence. The \textit{large $N$ limit} (where $N$ is usually the rank of the gauge group in the QFT) is that in which the expectation values of the multi-trace operators \textit{factorise} into those of the constituent single-trace operators. It is only in this limit that a QFT can have a holographic dual in the form of a classical gravity theory. Furthermore, when the QFT is strongly interacting, we expect there to be only few single-trace operators which have small scaling dimensions, because unless protected by symmetries there will be large quantum corrections to the anomalous dimensions at strong coupling. The remaining single-trace operators will decouple from the RG flow. The holographically dual classical gravity should then have only a few fields which are dual to the single-trace operators with small scaling dimensions.

Even in the large $N$ and strong coupling limit, the single-trace operators can mix with multi-trace operators along the RG flow \cite{Heemskerk:2010hk}. However, the RG flow can be thought of as a classical equations for scale evolution of single-trace operators in the sense that due to large $N$ factorisation, the multi-trace operators can be readily replaced by the products of the constituents single-trace operators when their expectation values are evaluated in \textit{any} state. It is expected that the gravitational theory can be truncated to pure gravity with a (negative) cosmological constant implying that there should be a consistent truncation of the dual RG flow equations to
\begin{equation}\label{schematic1}
\frac{\partial}{\partial \Lambda} t^\mu_{\phantom{\mu}\nu}(\Lambda) = F^\mu_{\phantom{\mu}\nu}[t^\mu_{\phantom{\mu}\nu}(\Lambda), \Lambda],
\end{equation}
with $F^\mu_{\phantom{\mu}\nu}$ being non-linear in $t^\mu_{\phantom{\mu}\nu}(\Lambda)$ so that it mixes with multi-trace operators built out of products of itself and it's derivatives along the RG flow.

In the strong interaction and large $N$ limits, it is then useful to conceive a RG flow such that (despite $t^\mu_{\phantom{\mu}\nu}(\Lambda)$ mixing with multi-trace operators constructed from products of itself and it's derivatives) at each scale there should exist an effective metric $g_{\mu\nu}(\Lambda)$ which is a non-linear functional of $t^\mu_{\phantom{\mu}\nu}(\Lambda)$ and $\Lambda$, i.e. of the form
\begin{equation}\label{schematic2}
g_{\mu\nu} (\Lambda) = g_{\mu\nu}[t^\mu_{\phantom{\mu}\nu}(\Lambda), \Lambda],
\end{equation}
which is constructed in the fixed background metric $\eta_{\mu\nu}$ such that $t^\mu_{\phantom{\mu}\nu}(\Lambda)$ preserves the form of the Ward identity
\begin{equation}\label{WILambda}
\nabla_{(\Lambda)\mu} t^\mu_{\phantom{\mu}\nu}(\Lambda) = 0,
\end{equation}
with $\nabla_{(\Lambda)}$ being the covariant derivative constructed from $g_{\mu\nu}(\Lambda)$. Therefore, an evolving metric $g_{\mu\nu}(\Lambda)$ which is a classical functional of $t^\mu_{\phantom{\mu}\nu}(\Lambda)$ (in the sense mentioned before) emerges as a tool for defining an efficient RG flow which invokes an efficient mixing of single-trace operators with multi-trace operators such that the Ward identity for local energy and momentum conservation takes the same form at each scale despite \textit{coarse-graining}. In order to find a dual field theory description we need to understand precisely how such a coarse-graining can be performed.  \textit{This property of preservation of form of Ward identity for local conservation of energy-momentum constitutes the major ingredient for defining an highly efficient RG flow.} This definition is not complete as it does not tell us how such a RG flow can be constructed in the field theory -- this will be described in the following section. Furthermore, we will also discuss the utility of such a RG flow in constructing strongly interacting large $N$ field theories. 

The major motivation of constructing a highly efficient RG flow is that it readily gives rise to a holographically dual classical gravity theory with full diffeomorphism invariance in one higher dimension due to the following theorem \cite{Behr:2015yna}.

\begin{theorem}
 Let us consider the $d-$dimensional scale evolution of $t^\mu_{\phantom{\mu}\nu}(\Lambda)$ taking the schematic form (\ref{schematic1}) in a \textit{fixed} background metric $g^{\rm (b)}_{\mu\nu}$ such that there exists a background metric $g_{\mu\nu}(\Lambda)$ which is a functional of $t^\mu_{\phantom{\mu}\nu}(\Lambda)$ and $\Lambda$ in the same \textit{fixed} background metric $g^{\rm (b)}_{\mu\nu}$ as schematically represented by (\ref{schematic2}), and in which $t^\mu_{\phantom{\mu}\nu}(\Lambda)$ satisfies the local conservation equation (\ref{WILambda}) at each $\Lambda$. Also let $g_{\mu\nu}(\Lambda)$ coincide with the fixed background metric $g^{\rm (b)}_{\mu\nu}$ at $\Lambda = \infty$ so that ${t^\mu_{\phantom{\mu}\nu}}^\infty$ satisfies $\nabla_{\rm(b)\mu}{t^\mu_{\phantom{\mu}\nu}}^\infty = 0$ with $\nabla_{\rm (b)}$ being the covariant derivative constructed from $g^{\rm (b)}_{\mu\nu}$.
 
We claim that as a consequence of the above assumptions, $g_{\mu\nu}(\Lambda)$ gives a $(d+1)-$dimensional bulk metric (\ref{FG}) in the Fefferman-Graham gauge with $r = \Lambda^{-1}$ such that it solves the equations of a \textit{pure} $(d+1)-$classical gravity theory with \textit{full} $(d+1)-$diffeomorphism invariance and a negative cosmological constant determined by the asymptotic curvature radius $l$. Also $g^{\rm (b)}_{\mu\nu}$ is the boundary metric of this emergent asymptotically AdS spacetime.
\end{theorem}
This theorem ensures that a $(d+1)-$dimensional classical gravity with full diffeomorphism invariance can be rewritten as a \textit{first order scale evolution} (\ref{schematic1}) of an effective energy-momentum tensor operator.

Let us now go back and see how Einstein's equation (\ref{Einstein}) can be reformulated into such a form as (\ref{schematic1}). Let us consider the background metric of the dual $4$ dimensional field theory to be $\eta_{\mu\nu}$ where the following RG flow equation \cite{Behr:2015yna}
\begin{eqnarray}\label{t-rg-example}
\frac{\partial t^\mu_{\phantom{\mu}\nu}(\Lambda)}{\partial \Lambda} &=& \frac{1}{\Lambda^3}\cdot\frac{1}{2} \Box t^{\mu}_{\phantom{\mu}\nu}(\Lambda)- \frac{1}{\Lambda^5}\cdot\left(\frac{1}{4}\, \delta^\mu_{\phantom{\mu}\nu}
{t^\alpha_{\phantom{\alpha}\beta}}(\Lambda)
{t^\beta_{\phantom{\beta}\alpha}}(\Lambda) - \frac{7}{128}\,\Box^2 {t^\mu_{\phantom{\mu}\nu}}(\Lambda)\right)
\nonumber\\&&+\frac{1}{\Lambda^5}\, \log\, \Lambda \cdot \frac{1}{32}\cdot \Box^2 {t^\mu_{\phantom{\mu}\nu}}(\Lambda)
+\mathcal{O}\left(\frac{1}{\Lambda^7}\, \log\, \Lambda\right)
\end{eqnarray}
can be constructed. For the above RG flow, we can indeed construct the following unique $g_{\mu\nu}(\Lambda)$ as given by
\begin{eqnarray}\label{g-example}
g_{\mu\nu}(\Lambda) &=& \eta_{\mu\nu} +\, \frac{1}{\Lambda^4}\cdot\frac{1}{4} \eta_{\mu\alpha}{t^\alpha_{\phantom{\alpha}\nu}}(\Lambda)
+\,\frac{1}{\Lambda^6}\cdot \frac{1}{24}\eta_{\mu\alpha}\Box {t^\alpha_{\phantom{\alpha}\nu}}(\Lambda)+
\nonumber\\&&
+ \frac{1}{\Lambda^8} \cdot \Bigg(\frac{1}{32}\,\eta_{\mu\alpha} {t^\alpha_{\phantom{\alpha}\rho}}(\Lambda)
{t^\rho_{\phantom{\rho}\nu}}(\Lambda) -\frac{7}{384}\, \eta_{\mu\nu}{t^\alpha_{\phantom{\alpha}\beta}}(\Lambda)
{t^\beta_{\phantom{\beta}\alpha}}(\Lambda)\nonumber\\&&\qquad\quad +\frac{11}{1536}\,\eta_{\mu\alpha}\Box^2 {t^\alpha_{\phantom{\alpha}\nu}}(\Lambda)\Bigg) +
\nonumber\\&& + \frac{1}{\Lambda^8}\,\log \, \Lambda\cdot \frac{1}{516} \cdot \eta_{\mu\alpha}\Box^2 {t^\alpha_{\phantom{\alpha}\nu}}(\Lambda)+ \mathcal{O}\left(\frac{1}{\Lambda^{10}}\, \log\, \Lambda\right)
\end{eqnarray}
as a functional of $t^\mu_{\phantom{\mu}\nu}(\Lambda)$ and $\Lambda$ in the flat Minkowski space background such that when it is considered as an effective background metric, the scale-dependent Ward identity (\ref{WILambda}) is satisfied at each $\Lambda$ (given that at $\Lambda= \infty$, the usual Ward identities (\ref{WIinfty}) hold). Furthermore, the $5-$dimensional bulk metric (\ref{FG}) then satisfies Einstein's equations (\ref{Einstein}) with $r = \Lambda^{-1}$ and the cosmological constant set to $-6/l^2$. The \textit{log} term in (\ref{t-rg-example}) is related to the conformal anomaly.

It is to be noted that the Ward identity (\ref{WILambda}) can also be recast as an effective operator equation, i.e. can be rewritten in a state-independent manner as an identity in flat Minkowski space $\eta_{\mu\nu}$. In the above example, (\ref{WILambda}) can be readily unpacked into
\begin{eqnarray}\label{opform}
\partial_\mu t^\mu_{\phantom{\mu}\nu}(\Lambda) &=& \frac{1}{\Lambda^4}\cdot\left(\frac{1}{16}\partial_\nu \left(t^\alpha_{\phantom{\alpha}\beta}(\Lambda)t^\beta_{\phantom{\beta}\alpha}(\Lambda)\right)-\frac{1}{8}t^\mu_{\phantom{\mu}\nu}(\Lambda)\partial_\mu\, {\rm Tr}\,t(\Lambda) \right)+\nonumber\\&&
+\frac{1}{\Lambda^6}\cdot\left(\frac{1}{48}t^\alpha_{\phantom{\alpha}\beta}(\Lambda)\partial_\nu\Box t^\beta_{\phantom{\beta}\alpha}(\Lambda)-\frac{1}{48}t^\mu_{\phantom{\mu}\nu}(\Lambda)\partial_\mu\Box\, {\rm Tr}\,t(\Lambda) \right)+\nonumber\\&&
+\mathcal{O}\left(\frac{1}{\Lambda^8}\right).
\end{eqnarray}
We then explicitly see that the scale-dependent effective background $g_{\mu\nu}(\Lambda)$ as given by (\ref{g-example}) serves to absorb the multi-trace contributions that spoil the usual Ward identity for local energy-momentum conservation. As a result, the Ward identity preserves its form (\ref{WILambda}) at each scale in the new scale-dependent background.

It should be immediately noted that although the RG flow (\ref{t-rg-example}) leads to the bulk metric in the Fefferman-Graham gauge, the classical gravity equations determining the latter should have underlying full diffeomorphism invariance. It can be readily argued that otherwise it is impossible that the RG flow (\ref{t-rg-example}) will be able to preserve a Ward identity of the form (\ref{WILambda}). In particular, absence of diffeomorphism invariance in the dual bulk theory that gives the evolution of $g_{\mu\nu}(\Lambda)$ will imply that there will be other propagating degrees of freedom in addition to $g_{\mu\nu}(\Lambda)$ in which case the Ward identity (\ref{WILambda}) should be modified. 

The RG flow reformulation (\ref{t-rg-example}) of Einstein's equations has been demonstrated so far only in the asymptotic (i.e. UV) expansion. This series (\ref{t-rg-example}) has a \textit{finite} radius of convergence related to the scale (radius) where the Fefferman-Graham coordinates has a coordinate singularity in the dual spacetime. In order to sum  (\ref{t-rg-example}) to all orders in $\Lambda^{-1}$, we need to assume a specific form of the energy-momentum tensor such as the hydrodynamic form to be considered later. In the latter case, all orders in $\Lambda^{-1}$ can be summed at any given order in derivative expansion. The radius of convergence is the scale corresponding to the location of the horizon at late time and is related to the final temperature.

The immediate question is how do we derive the RG flow reformulation of the classical gravity equations such as (\ref{t-rg-example}) corresponding to Einstein's equations. In order to answer this, it is sufficient to understand what does $t^\mu_{\phantom{\mu}\nu}(\Lambda)$ correspond to in the dual gravitational theory. To do this a gauge-independent formulation of the map between RG flow and gravitational equations is helpful. We express the $(d+1)-$dimensional spacetime metric via ADM-like variables \cite{Arnowitt:1962hi}:
\begin{equation}\label{ADM}
{\rm d}s^2 = \alpha(r,x) {\rm d}r^2 + \gamma_{\mu\nu}(r,x)\left({\rm d}x^\mu + \beta^\mu(r,x) {\rm d}r\right)\left({\rm d}x^\nu + \beta^\nu(r,x) {\rm d}r\right).
\end{equation}
in which $\alpha$ is the analogue of the lapse function and $\beta^\mu$ is the analogue of the shift vector. Specifying conditions determining these amounts to gauge-fixing the diffeomorphism symmetry. For reasons (state independence and absence of explicit presence of $l$ in the evolution equations) mentioned before, the identification of $\Lambda$ and $g_{\mu\nu}(\Lambda)$ should take the following forms assuming that $r=0$ is the boundary \cite{Behr:2015yna}:
\begin{equation}
r = \Lambda^{-1}, \quad g_{\mu\nu}(\Lambda = r^{-1}) = \frac{r^2}{l^2}\gamma_{\mu\nu}(r,x).
\end{equation}
Note the above is not only true for Einstein's gravity but also for a general gravitational theory. In this case the form of $t^\mu_{\phantom{\mu}\nu}(\Lambda)$ can also be fixed up to an overall multiplicative constant by (i) requiring it to be state (solution) independent, (ii) demanding absence of explicit presence of $l$ in its scale evolution, and (iii) requiring that it satisfies the Ward identity (\ref{WILambda}). In a general gravitational theory, these imply that $t^\mu_{\phantom{\mu}\nu}(\Lambda)$ should take the form \cite{Behr:2015yna}:
\begin{equation}\label{tandT}
t^\mu_{\phantom{\mu}\nu}(\Lambda = r^{-1}) =\left(\frac{l}{r}\right)^d\cdot\left({T^\mu_{\phantom{\mu}\nu}}^{\rm ql} +{T^\mu_{\phantom{\mu}\nu}}^{\rm ct}\right),
\end{equation}
up to an overall multiplicative constant, where ${T^\mu_{\phantom{\mu}\nu}}^{\rm ql}$ is the quasi-local stress tensor that is conserved via equations of motion \cite{Balcerzak:2007da} and ${T^\mu_{\phantom{\mu}\nu}}^{\rm ct}$ is a sum of gravitational counterterms built out of the Riemann curvature of $\gamma_{\mu\nu}$ and its covariant derivatives such that they satisfy (\ref{WILambda}) via Bianchi-type identities. Up to second order in derivatives, ${T^\mu_{\phantom{\mu}\nu}}^{\rm ct}$ can be parametrised as:
\begin{eqnarray}\label{Tct}
{T^\mu_{\phantom{\mu}\nu}}^{\rm ct} &=&-\frac{1}{8 \pi G_N}\Bigg[C_{(0)}\cdot \frac{1}{l} \cdot \delta^\mu_{\phantom{\mu}\nu} + C_{(2)}\cdot l \cdot \left(R^\mu_{\phantom{\mu}\nu}[\gamma] - \frac{1}{2}R[\gamma]\delta^\mu_{\phantom{\mu}\nu}\right) +
 \cdots\,\Bigg],
\end{eqnarray}
with $C_{(n)}$s being dimensionless constants that depend on the gravitational theory and $G_N$ being the $(d+1)-$dimensional gravitational constant. Above, the indices have been lowered/raised by the induced metric $\gamma_{\mu\nu}$/its inverse. In the case of Einstein's gravity, ${T^\mu_{\phantom{\mu}\nu}}^{\rm ql}$ is the Brown-York tensor:
\begin{equation}\label{Brown-York}
{T^\mu_{\phantom{\mu}\nu}}^{{\rm ql}} = -\frac{1}{8 \pi G_N} \, \gamma^{\mu\rho}\left(K_{\rho\nu}- K \gamma_{\rho\nu}\right).
\end{equation}
Here $K_{\mu\nu}$ is the extrinsic curvature of the hypersurface $r = \textit{constant}$ given by
\begin{equation}\label{extrinsic-curvature}
K_{\mu\nu} = -\frac{1}{2\alpha}\left(\frac{\partial \gamma_{\mu\nu}}{\partial r} - \nabla_{(\gamma)\mu} \beta_\nu -\nabla_{(\gamma)\nu} \beta_\mu \right),
\end{equation}
with $\beta_\rho = \gamma_{\rho\mu} \beta^\mu$, and $K = K_{\mu\nu}\gamma^{\mu\nu}$. Therefore, in the Fefferman-Graham gauge, $t^\mu_{\phantom{\mu}\nu}(\Lambda)$ should take the following form for Einstein's gravity:
\begin{eqnarray}\label{tFG}
t^\mu_{\phantom{\mu}\nu}(\Lambda= r^{-1}) &=& \frac{l^{d-1}}{16\pi G_N}\Bigg[\frac{1}{r^{d-1}}\cdot \left(z^\mu_{\phantom{\mu}\nu} - ({\rm Tr}\, z) \,\delta^\mu_{\phantom{\mu}\nu}\right) +2\cdot \frac{1}{r^d}\cdot  \left( d- 1 - C_{(0)}\right) \cdot \delta^\mu_{\phantom{\mu}\nu}-\nonumber\\&&
\quad\quad\quad\quad - 2\cdot \frac{1}{r^{d-2}}\cdot C_{(2)}\cdot \left(R^\mu_{\phantom{\mu}\nu}[g] - \frac{1}{2}R[g]\delta^\mu_{\phantom{\mu}\nu}\right)
+ \cdots \Bigg].
\end{eqnarray}
The overall multiplicative constant $l^{d-1}/(16\pi G_N)$ has been chosen by us above and cannot be fixed by the arguments presented before. This overall factor is actually proportional to $N^2$ of the dual field theory (as mentioned before $l$ itself has no meaning in the dual QFT but the gravitational constant measured in units where $l=1$ does have one). This overall factor can be fixed by identifying the temperature in the field theory in a thermal state to that of the Hawking temperature of the dual black hole. This however requires taking into account quantum effects. For later convenience, we rescale $t^\mu_{\phantom{\mu}\nu}(\Lambda)$ by this overall factor $(16\pi G_N)/l^{d-1}$ so that $N^2$ is now absorbed in the definition of $t^\mu_{\phantom{\mu}\nu}(\Lambda)$. There is still a genuine ambiguity in the definition of $t^\mu_{\phantom{\mu}\nu}(\Lambda)$ which arises from the choice of the gravitational counterterm coefficients $C_{(n)}$s. Fixing this ambiguity leads us to a profound and surprising understanding of gravity itself as described below.

We first observe that the above ambiguity of choosing coefficients of gravitational counterterms has an immediate consequence for the map between gravity and RG flow. It implies that the equations of gravity can be reformulated into infinitely many RG flow equations of the form (\ref{schematic1}) for any choice of gauge fixing of bulk diffeomorphisms. Each of these formulations corresponds to a specific choice of gravitational counterterms $C_{(n)}$s. Furthermore, each such RG flow will require the existence of the same (unique) $g_{\mu\nu}(\Lambda)$ taking the schematic form (\ref{schematic2}) in which the effective Ward identity (\ref{WILambda}) will be satisfied, and which will lead to the same bulk metric that satisfies the dual diffeomorphism invariant gravitational equations with a specific gauge fixing. 

It is of course desirable that at the UV fixed point, i.e. at $\Lambda = \infty$, ${t^\mu_{\phantom{\mu}\nu}}^\infty$ is finite. This leads to fixing a finite number of leading counterterms, particularly \cite{Henningson:1998ey,Balasubramanian:1999re,deBoer:1999tgo}
\begin{equation}\label{C0C2}
C_{(0)} = d -1, \quad C_{(2)} = - \frac{1}{d-2}, \quad \text{etc.}
\end{equation}
It is interesting to note that ${t^\mu_{\phantom{\mu}\nu}}^\infty$ is completely free of ambiguities when the boundary metric is $\eta_{\mu\nu}$, because all other counterterms, except a few leading terms, vanish in any asymptotically AdS space because of the enhancement of symmetries in the geometry in the asymptotic limit. We thus recover the result for ${t^\mu_{\phantom{\mu}\nu}}^\infty$ as in the traditional AdS/CFT correspondence. This procedure is however unsatisfactory for two reasons. Firstly, we still have infinite ambiguities in the form of unfixed coefficients of the infinite number of gravitational counterterms which vanish asymptotically. Secondly, if we can genuinely rewrite gravity as RG flow,  in the latter form it should be first order evolution so that we can either specify conditions at the UV or at the IR, but not at both places. It is more desirable that we restrict the IR as we need a sensible IR behaviour of the RG flow even in cases where the UV completion is unknown. This is specially relevant for finding holographic duals of theories like QCD where only the IR can be expected to be captured by a holographically dual classical gravity description at large $N$ \cite{Klebanov:2000hb} -- in the UV the emergent geometry can have a singularity implying the necessity of new degrees of freedom.

This ambiguity is fixed by the following theorem stated below \cite{Kuperstein:2013hqa,Behr:2015yna,Behr:2015aat}.
\begin{theorem}
Up to an overall multiplicative constant for $t^\mu_{\phantom{\mu}\nu}(\Lambda)$, there is a unique choice of the functional $F^\mu_{\phantom{\mu}\nu}$ in (\ref{schematic1}) that reformulates a pure holographic classical gravity theory as RG flow such that the endpoint of the RG flow at $\Lambda = \Lambda_{\rm IR}$ can be converted to a fixed point in the hydrodynamic limit corresponding to \textit{non-relativistic incompressible Navier-Stokes fluid} under the universal rescaling:
\begin{equation}\label{rescale}
\Lambda_{\rm IR}^{-1} - \Lambda^{-1}  = \xi \cdot \lambda^{-1} \quad t = \frac{\tau}{\xi},
\end{equation}
(corresponding to near horizon and long time behaviour of the dual gravitational dynamics) where $\xi$ is taken to zero with $\lambda$ and $\tau$ kept finite. This also corresponds to fixing the gravitational counterterms in (\ref{Tct}) uniquely so that $t^\mu_{\phantom{\mu}\nu}(\Lambda)$ is uniquely identified as a functional of the ADM variables in the dual pure gravitational theory. Even those counterterms which are necessary to cancel UV divergences are also determined by the prescribed IR behaviour.
\end{theorem}
Remarkably, the hydrodynamic limit can fix all the ambiguities of the RG flow which however has a state-independent formulation in terms of evolution of the operator $t^\mu_{\phantom{\mu}\nu}(\Lambda)$ with the scale and which is valid even beyond this limit. Thus long wavelength perturbations of black holes unsurprisingly play a very fundamental role in understanding holographic correspondence as RG flow. We do not have a complete proof of this theorem, so actually it is still a conjecture. However very non-trivial calculations which will be sketched in the next section provide solid supporting verifications.

It is also important that the end point of the RG flow is not really fixed point although it becomes so after the rescaling (\ref{rescale}) which has been first introduced in the context of gravitational dynamics in the hydrodynamic limit in the dual theory \cite{Bredberg:2010ky}. As we will see in the next section, it implies that all physical parameters in $t^\mu_{\phantom{\mu}\nu}(\Lambda)$ should satisfy appropriate bounds regrading how they behave at the endpoint. These bounds determine all integration constants of the first order RG flow and thus determine the UV values of physical observables. Remarkably, these UV values are exactly the same as those for which dual gravitational geometries are free of naked singularities. Since the hydrodynamic limit determines the RG flow uniquely, all physical observables beyond the hydrodynamic limit can also be obtained from the RG flow. \textit{Therefore, not only that a holographic gravitational theory can be reformulated as a unique RG flow for every choice of gauge-fixing of diffeomorphism symmetry (up to an overall constant numerical factor for $t^\mu_{\phantom{\mu}\nu}(\Lambda)$) , the data which leads to regular horizons are also determined by this RG flow.} This IR criterion constitutes another crucial defining feature of a highly efficient RG flow as exemplified by (\ref{t-rg-example}) for Einstein's gravity. In the following section, we will present more details on how this IR criterion fixes the ambiguity of gravitational couunterterms leading to a unique highly efficient RG flow for each choice of gauge-fixing of diffeomorphism symmetry in the dual classical gravity equations.

Finally, we note that the choice of gauge fixing of the diffeomorphism symmetry is also encoded in the RG flow (which in cases other than the Fefferman-Graham gauge may contain auxiliary non-dynamical variables corresponding to the lapse function and the shift vector). This is due to the feature that any asymptotically AdS metric has a residual gauge symmetry which corresponds to conformal transformations for the dual theory at the boundary under which the dual theory must be invariant (up to quantum anomalies that are related to logarithmic terms necessary for regulating divergences of the on-shell gravitational action \cite{Henningson:1998ey,Balasubramanian:1999re}). Such diffeomorphisms which preserve the Fefferman-Graham gauge are called Penrose-Brown-Henneaux (PBH) transformations in the literature \cite{Penrose,Brown:1986nw,Schwimmer:2000cu}, and these can be readily generalised to other choices of gauge fixing \cite{Behr:2015yna}. These turn out to lead to automorphism symmetry of the dual RG flow equations (\ref{schematic1}) when they are formulated in a general fixed conformally flat background metric \cite{Behr:2015yna}. We have called this \textit{lifted Weyl symmetry}. Deciphering this symmetry for a given highly efficient RG flow readily leads us to determine the corresponding gauge fixing in the dual gravity theory and thus also the choice of hypersurface foliation in the dual geometries used as holographic screens at various scales. 

\section{The field theory perspective and the hydrodynamic limit}
In the previous section, we have discussed reformulation of a holographic pure gravity theory as a highly efficient RG flow which can self-determine microscopic UV data by an appropriate IR criterion, and reproduce results of traditional holographic correspondence where these data are determined by explicitly solving the gravitational equations and demanding absence of naked singularities. In this section, following \cite{Behr:2015aat} we will show how such a RG flow can be constructed in the field theory and even define it constructively in the strong interaction and large $N$ limits. We will illustrate the construction briefly in the hydrodynamic limit.

In the strong interaction and large $N$ limits, a handful of single-trace operators (dual to the fields in the gravitational theory) can define at least some sectors of the full theory in the sense mentioned in the previous section. Instead of using the elementary fields to define the QFT, it then makes sense to use collective variables which are directly measurable and which parametrise the expectation values of these single-trace operators in all states. Such collective variables include the hydrodynamic variables and can be extended to include the shear-stress tensor and other non-hydrodynamic parameters also (see for innstance \cite{Iyer:2009in,Iyer:2011qc,Heller:2014wfa}). At the very outset, it is clear that such an exercise of defining quantum operators via collective variables which parametrise their expectation values is futile except in the strong interaction and large $N$ limits. Unless we are in the large $N$ limit, the expectation values of the multi-trace operators do not factorise, therefore we need new collective variables for defining multi-trace operators. Also if we are not in the strong interaction limit, we will need to consider infinitely many single-trace operators. These will imply proliferation of the number of collective variables required to describe exact quantum dynamics.

The physical picture is as follows. Consider a set of microscopic single-trace operators $O_I^\infty$ such as the energy-momentum tensor which can be parametrised by a set of collective variables $X_A^\infty$ such as the hydrodynamic variables. Furthermore, the spacetime evolution of the expectation values $\langle O_I^\infty \rangle$ can be captured by equations of motions for the collective variables $X_A^\infty$ such as the hydrodynamic equations with parameters $\eta_M^\infty$ such as the transport coefficients. It is to be noted here that the hydrodynamics being mentioned here is not referring to any kind of coarse-graining, rather an asymptotic series involving perturbative derivative expansion (with infinite number of transport coefficients) which captures the dynamics near thermal equilibrium \cite{Heller:2013fn,Basar:2015ava}. Generally speaking, we can succinctly represent the quantum operators $O_I^\infty$ through their expectation values $\langle O_I^\infty \rangle[X_A^\infty,\eta_M^\infty ]$.

We can readily do an appropriate coarse-graining of our measurements of $\langle O_I^\infty \rangle$ and proceed to define $\langle O_I (\Lambda) \rangle$. The latter definition can be achieved via appropriate coarse-grained collective variables $X_A(\Lambda)$ which by construction follow similar equations as $X_A(\infty)$ but with new parameters $\eta_M(\Lambda)$. As in any RG flow, we expect that we need fewer parameters $\eta_M(\Lambda)$ to describe the spacetime evolution of $X_A(\Lambda)$ than the number of $\eta_M^\infty$ we need to describe that of $X_A^\infty$ to the same degree of approximation. In a highly efficient RG flow, we define the coarse-grained quantum operators $O_I(\Lambda)$ through their expectation values $\langle O_I(\Lambda) \rangle[X_A(\Lambda),\eta_M(\Lambda) ]$ assuming that the coarse-grained operators are the same functionals of the coarse-grained collective variables at each scale (as in the UV) but with new scale-dependent parameters. Note that there is no explicit dependence on $\Lambda$ in the functionals $\langle O_I(\Lambda) \rangle[X_A(\Lambda),\eta_M(\Lambda) ]$.

In order to complete the construction we will need to define the constructive principles for coarse-graining that defines $X_A(\Lambda)$ which should follow similar equations at each scale but with new scale-dependent parameters $\eta_M(\Lambda)$. These three principles are listed below.
\begin{enumerate}
\item \textbf{High efficiency:}\\There should exist an appropriate background metric: \\$g_{\mu\nu}(\Lambda)[X_A(\Lambda), \eta_M(\Lambda), \Lambda]$\\and appropriate background sources:\\ $J(\Lambda)[X_A(\Lambda), \eta_M(\Lambda), \Lambda]$ \\ at each $\Lambda$ such that the Ward identity
\begin{equation}\label{WILambdanew}
\nabla_{(\Lambda)\mu}t^\mu_{\phantom{\mu}\nu}(\Lambda) = {\sum}' O_I(\Lambda) \nabla_{(\Lambda)\nu}J_I(\Lambda)
\end{equation} 
is satisfied with $\nabla_{(\Lambda)}$ being the covariant derivative constructed from $g_{\mu\nu}(\Lambda)$ and $\sum'$ denoting summation over all effective single-trace operators except $t^\mu_{\phantom{\mu}\nu}(\Lambda)$.\\
\item \textbf{Upliftability to operator dynamics:} \\The functionals $g_{\mu\nu}(\Lambda)[X_A(\Lambda), \eta_M(\Lambda), \Lambda]$ and $J(\Lambda)[X_A(\Lambda), \eta_M(\Lambda), \Lambda]$ can be uplifted to functionals of the single-trace operators. Therefore, they should assume the forms\\
$g_{\mu\nu}(\Lambda)[O_I(\Lambda),\Lambda]$ and $J(\Lambda)[O_I(\Lambda),\Lambda]$ \\
so that the effective Ward identities (\ref{WILambdanew}) can be promoted to operator equations such as (\ref{opform}).  As a consequence, it follows that the scale evolution equations for $O_I(\Lambda)$ such as (\ref{t-rg-example}) become state-independent equations involving single and multi-trace operators and $\Lambda$ only, and thus without involving the collective variables explicitly.\\
\item \textbf{Good endpoint behaviour:}\\
 The IR end point of the RG flow where most of the parameters $\eta_M(\Lambda)$ blow up and some collective variables $X_A(\Lambda)$ become singular can be made regular under the universal rescaling (\ref{rescale}) corresponding to near horizon and long time limits of the dual spacetimes. In the hydrodynamic limit, the endpoint should be converted to a fixed point corresponding to non-relativistic incompressible Navier-Stokes equations under the stated rescaling.
\end{enumerate}

Our claim is that for every realisation of a highly efficient RG flow which satisfies the above three principles:
\begin{enumerate}
\item there corresponds a unique dual gravitational theory up to a choice of gauge-fixing of the bulk diffeomorphism symmetry that can have a dual holographic description as a strongly interacting large $N$ QFT, and
\item there is unique set of UV data for (the infinitely many) $\eta_M(\Lambda)$ which however can be resummed in the IR so that the dynamics at the endpoint can be described by a finite number of parameters (such as the shear viscosity of the infrared non-relativistic incompressible Navier-Stokes fluid), and also these UV data (such as the UV values of the infinitely many transport coefficients) are the same as those which lead to the regularity of the future horizons in the dual gravitational theory corresponding to the RG flow.
\end{enumerate}
The infrared end point typically corresponds to the location of the horizon at late time, and thus the highly efficient RG flow connects the AdS/CFT correspondence with the membrane paradigm. The highly efficient RG flow gives a constructive way to define strongly interacting large $N$ QFTs by reformulating the holographic correspondence. The first two principles in our list defining highly efficient RG flows utilise the first theorem of reformulation of diffeomorphism invariant gravity and the third  principle in our list utilises the second theorem discussed in the previous section. However, here our list of principles also presents a generalisation which is valid not only for the reconstruction of holographic pure gravity but also when the latter is coupled to a finite number of matter fields. The utility of highly efficient RG flow is actually deeper. It shows that all such QFTs and hence all holographic gravitational theories are determined by finite amount of data that governs the dynamics at the end point. Therefore, all holographic gravitational theories can be parametrised by a finite number of free parameters in any given dimension. How this parametrisation works has not been completely understood yet.

As an illustration, let us see how we construct highly efficient RG flows in the hydrodynamic limit \cite{Behr:2015aat}. Once again, let us revert back to the sector of states where $t^\mu_{\phantom{\mu}\nu}(\Lambda)$ is the only single-trace operator with a non-vanishing expectation value for the sake of simplicity. The expectation value of $t^\mu_{\phantom{\mu}\nu}(\Lambda)$ is parametrised by the (collective) hydrodynamic variables $u^\mu(\Lambda)$ and $T(\Lambda)$ which thus define the quantum operator. Furthermore, $u^\mu(\Lambda)$ can be assumed to satisfy Landau-Lifshitz definition in which case $u^\mu(\Lambda)$ is a timelike eigenvector of $t^\mu_{\phantom{\mu}\nu}(\Lambda)$ with unit norm with respect to the background metric $g_{\mu\nu}(\Lambda)$ so that $u^\mu(\Lambda) g_{\mu\nu}(\Lambda)u^\nu(\Lambda) =-1$. The hydrodynamic variables $u^\mu(\Lambda)$ and $T(\Lambda)$ should satisfy hydrodynamic equations in the effective background $g_{\mu\nu}(\Lambda)$ with scale-dependent energy density $\epsilon(\Lambda)$, pressure $P(\Lambda)$ and transport coefficients $\gamma^{(n,m)}(\Lambda)$, where $n$ denotes the order in the derivative expansion (running from zero to infinity) and $m$ lists the finite number of independent parameters at each order in the derivative expansion. At the first order in the derivative expansion, there are only two independent transport coefficients, namely the shear and the bulk viscosities. 

The coarse-graining of $u^\mu(\Lambda)$ and $T(\Lambda)$ can be expressed both in integral or differential form. The latter form is more useful and is as shown below:
\begin{eqnarray}\label{uTevol}
:\frac{\partial u^{\mu} (\Lambda)}{\partial \Lambda}: &=& a^{(0)}(\Lambda) u^\mu(\Lambda) + \sum_{n=1}^\infty \sum_{m=1}^{n_{\rm s}}a^{(n,m)}_{\rm s}(\Lambda)\, \mathcal{S}^{(n,m)}(\Lambda)\, u^\mu(\Lambda) +\nonumber\\&&+\sum_{n=1}^\infty \sum_{m=1}^{n_{\rm v}}a^{(n,m)}_{\rm v}(\Lambda) \,{\mathcal{V}^\mu}^{(n,m)}(\Lambda)\, , \nonumber\\
:\frac{\partial T (\Lambda)}{\partial\Lambda}: &=& b^{(0)}(\Lambda)  + \sum_{n=1}^\infty \sum_{m=1}^{n_{\rm s}}b^{(n,m)}_{\rm s}(\Lambda) \mathcal{S}^{(n,m)}(\Lambda).
\end{eqnarray}
Above $\mathcal{S}^{(n,m)}$ denotes the independent hydrodynamic scalars that can be constructed from derivatives of $u^\mu(\Lambda)$ and $T(\Lambda)$ at the $n-th$ order in derivatives (with independent meaning that a linear sum of these scalars do not vanish using lower order equations of motion). When $n=1$, there is only one such scalar, namely $(\partial\cdot u)$. Similarly, ${\mathcal{V}^\mu}^{(n,m)}(\Lambda)$ denotes hydrodynamic vectors which are not parallel to $u^\mu(\Lambda)$ (as otherwise it can be expressed via a scalar multiplying $u^\mu(\Lambda)$). When $n=1$, there is only one such vector, namely $(u(\Lambda)\cdot \partial)u^\mu(\Lambda)$.
The symbols $:\cdots:$ stand for subtracting away non-hydrodynamic contributions. The coarse-graining actually arises from a truncation of the series (\ref{uTevol}) at a given order in the derivative expansion. So far this is the most general way to coarse-grain hydrodynamic variables which is consistent with the hydrodynamic limit.

Furthermore, we assume that the flow of the energy density, pressure and the transport coefficients take the form of ordinary differential equations:
\begin{eqnarray}\label{transODE}
\frac{{\rm d}\epsilon(\Lambda)}{{\rm d}\Lambda} &=& K[\epsilon(\Lambda), P(\Lambda), \Lambda], \nonumber\\
\frac{{\rm d}P(\Lambda)}{{\rm d}\Lambda} &=& L[\epsilon(\Lambda), P(\Lambda), \Lambda],\nonumber\\
\frac{{\rm d}\gamma^{(n,m)}(\Lambda)}{{\rm d}\Lambda} &=& M^{(n,m)}[\epsilon(\Lambda), P(\Lambda), \gamma^{(k \leq n, p)}(\Lambda), \Lambda],
\end{eqnarray} 
in which the scale evolution of transport coefficients at $n-$th order in derivative expansion involves only those at the same or lower orders. 

The mathematical problem of constructing highly efficient RG flows in the hydrodynamic limit now becomes well-defined. We simply need to solve for the parameters $a^{(0)}$, $b^{(0)}$, $a^{(n,m)}_{\rm s}$, $a^{(n,m)}_{\rm v}$, $b^{(n,m)}_{\rm s}$ in (\ref{uTevol}) and the functionals $K$, $L$ and $M^{(n,m)}$ appearing in (\ref{transODE}) such that the three principles listed before are satisfied. Unfortunately, we do not yet know how this mathematical problem can be solved directly. Fortunately, there is a concrete algorithmic method \cite{Kuperstein:2013hqa} (developed using some results of \cite{Gupta:2008th,Kuperstein:2011fn}) to reformulate the classical gravitational equations in the forms (\ref{uTevol}) and (\ref{transODE}) which can be used to solve for these parameters indirectly so that we can satisfy the three principles and obtain all highly efficient RG flows.

The most subtle aspect of this procedure is in how we satisfy the third principle of good infrared behaviour. As discussed in the previous section, the reformulation of gravity as RG flow is subject to the ambiguities of undetermined counterterm coefficients as presented in (\ref{Tct}) before. However there are a finite number of such terms at each order in the derivative expansion. The recipe is to proceed with these ambiguities which lead to unknown numerical constants in (\ref{uTevol}) and (\ref{transODE}). In order for the endpoint to be governed by non-relativistic incompressible Navier-Stokes equations, $\epsilon(\Lambda)$ must be finite at the endpoint $\Lambda_{\rm IR}$ where $\gamma^{(n,m)}(\Lambda)$ should satisfy bounds $\gamma^{(n,m)}(\Lambda)\leq (\Lambda -\Lambda_{\rm IR})^{-k(n,m)}$ with $k(n,m)$ being appropriate numerical constants which are independent of the RG flow or the dual gravitational theory \cite{Kuperstein:2013hqa}. It turns out that when we actually solve for $\gamma^{(n,m)}(\Lambda)$ the number of terms which diverge worse than the prescribed bounds are typically more than the number of integration constants available unless the counterterm coefficients which have been left undetermined so far are precisely chosen at each order in the derivative expansion. Setting these counterterm coefficients to such values, we can fix all integration constants of the RG flow and thus we can determine the UV values of all transport coefficients \textit{uniquely}.

This procedure has been explicitly implemented for Einstein's gravity at zeroth, first and second orders in the derivative expansion. Remarkably, the UV values of the equations of state and the first and second order transport coefficients determined via this method matches exactly with the known values \cite{Policastro:2001yc,Baier:2007ix,Bhattacharyya:2008jc} which are required for the regularity of the future horizon. The methods of explicit construction of highly efficient RG flows can be generalised beyond the hydrodynamic limit by including non-hydrodynamic collective variables \cite{Iyer:2009in,Iyer:2011qc,Iyer:2011ak,Heller:2013fn,Heller:2014wfa,Basar:2015ava} as discussed in the literature before.

We should understand how to solve for highly efficient RG flows independently without using the theorems for reformulation of dual gravitational theories so that we can classify all gravitational theories that are holographic and also where a finite number of IR parameters can determine all microscopic UV data in the dual theories.

\section{Outlook}
We have demonstrated that the reformulation of classical gravity as RG flow not only reveals how the holographic duality works but also gives us a deeper understanding of gravitational dynamics itself, in particular relating to what kind of data that determine the spacetime metric lead to absence of naked singularities. 

An outstanding issue is to take another step to understand how to include quantum corrections in gravity while mapping it to a highly efficient RG flow whose notion also needs to be further generalised to go beyond the large $N$ limit. In order to proceed, it should be useful to understand better how the three principles which define highly efficient RG flows themselves originate from a simpler and more holistic principle. Such a direction seems possible as there is evidence that classical gravity emerges from features of quantum entanglement in dual quantum systems \cite{Faulkner:2013ica}. In particular, it is known that classical minimal surfaces in dual geometries encode entanglement entropies in dual field theories \cite{Ryu:2006bv}. It has also been argued elsewhere that efficient nonperturbative RG flows that coarse-grain quantum information efficiently such that they remove short range entanglement but preserve long range entanglement give rise to the holographic correspondence \cite{PhysRevD.86.065007}. It is natural to speculate that when quantum gravity corrections are included the infrared end point for the dual RG flow is not characterised necessarily by local order parameters, but rather by non-local quantum order parameters related to patterns of global long range entanglement. This point of view also has a potential for defining quantum geometry in the emergent gravity theory.

We hold the point of view that a breakthrough in this direction is likely to come from a reformulation of classical gravity equations themselves which uses non-local geometric objects such as geodesics and minimal surfaces as the dynamical variables, and also which makes a tangible connection with the local RG flow perspective described in the present article. At present, how this can be realised seems a bit mysterious, however  it is very likely that there are hidden treasures in classical gravity which are yet to discovered. It will not be surprising if the surface terms\cite{Mukhopadhyay:2006vu,Padmanabhan:2007en} introduced by T. Padmanbhan, and his novel variational principle involving these surface terms which give classical gravitational equations in the bulk without using the metric as a dynamical variable, can shed some light in this direction. Another interesting reformulation \cite{deBoer:2016pqk} of classical gravity equations involving objects which are analogous to minimal surfaces but also sensitive to the operator content in the dual QFTs has appeared recently.

Finally, I would like to mention that the reformulation of classical gravity equations as RG flows has also informed the development of a new approach for combining weak and strong coupling degrees of freedom of the quark-gluon plasma produced by heavy ion collisions self-consistently into a novel nonperturbative framework \cite{Iancu:2014ava,Mukhopadhyay:2015smb}. Unravelling the holographic origin of gravity will surely revolutionise our understanding of nonperturbative aspects of quantum dynamics in the future.
\section*{Acknowledgments}
The research of A.M. is supported by a Lise-Meitner fellowship
of the Austrian Science Fund (FWF), project no. M 1893-N27. A part of this article has been used as a contribution for the festschrift in honour of the 60th anniversary celebration of Prof. T. Padmanabhan.

\bibliographystyle{utphys}
\bibliography{paddyat60ayan}
\end{document}